\documentclass[final,5p,times,twocolumn]{elsarticle}

\usepackage{graphics}

\usepackage{amssymb}

\journal{Nuclear Instruments and Methods A}

\begin{document}

\begin{frontmatter}



\title{A neutron spectrometer for studying giant resonances with $(p,n)$ reactions in inverse kinematics}


\author[label1]{L. Stuhl}
\address[label1]{ATOMKI, Institute for Nuclear Research, Hungarian Academy of Sciences, Debrecen, Hungary}
\ead{stuhl@atomki.hu}
\author[label1]{A. Krasznahorkay}
\author[label1]{M. Csatl\'os}
\author[label1,label2]{A. Algora}
\address[label2]{Instituto de Fisica Corpuscular, Univ. Valencia - C.S.I.C. Edificio de Institutos de Paterna, Valencia, Spain}
\author[label1]{J. Guly\'as}
\author[label1]{G. Kalinka}
\author[label1]{J. Tim\'ar}
\author[label3]{N. Kalantar-Nayestanaki}
\address[label3]{Kernfysisch Versneller Instituut, University of Groningen, Groningen, The Netherlands}
\author[label3]{C.~Rigollet}
\author[label3]{S.~Bagchi}
\author[label3]{and M. A. Najafi}

\begin{abstract}
\indent
A neutron spectrometer, the European Low-Energy Neutron Spectrometer (ELENS), has been constructed to study exotic nuclei in inverse-kinematics experiments. The spectrometer, which consists of plastic scintillator bars, can be operated in the neutron energy range of 100 keV to 10 MeV. The neutron energy is determined using the time-of-flight technique, while the position of the neutron detection is deduced from the time-difference information from photomultipliers attached to both ends of each bar. A novel wrapping method has been developed for the plastic scintillators. The array has a larger than 25\% detection efficiency for neutrons of approximately 500 keV in kinetic energy and an angular resolution of less than 1 degree. Details of the design, construction and experimental tests of the spectrometer will be presented.
\end{abstract}


\end{frontmatter}


\section{Introduction}
\indent
Nuclear-structure studies are clearly shifting toward the isotopes that are far from the valley of stability. In earlier nuclear-physics studies, the nuclear reactions induced by light charged particles (such as $^1$H, $^2$H, $^3$He and $^4$He) turned out to be very useful for studying nuclear structure. Recently, these reactions have also begun to be used in radioactive beams in inverse kinematics. The use of radioactive beams and inverse kinematics often requires very special targets (gas, liquid, etc.). Therefore, detection systems that are optimized for these experimental conditions are required. The information of interest is then extracted from the kinematical characteristics of the reaction products, such as their scattering angles and energies. The production rate of exotic nuclei decreases exponentially with the increase in the proton-neutron asymmetry \cite{tani}. To counterbalance this effect, high-efficiency detector setups and thick reaction targets are required, which lead to large uncertainties in the kinematic reconstruction and energy resolution \cite{mittig}.

\indent
The detection of a low-energy recoil product is often rather difficult, especially in the case of L=0 transitions, in which the cross section peaks at approximately $0^{\circ}$, and one must detect the light reaction products at small angles and with energies that usually extend below 1 MeV. Various experimental methods have been developed to overcome the above difficulties. One approach to studying such reactions is to detect the ejected 
low-energy charged particles in active gas targets \cite{cam}, \cite{tanihata}, \cite{roger}. Moreover, it is often desirable to detect reaction products in coincidence with each other to enhance the selection of the studied reaction channels and reduce the background contribution. The second option is to use neutrons as the outgoing particles. In contrast to charged reaction products, neutrons can pass through relatively thick materials without scattering and losing their energy, so in these experiments, relatively thick targets can be utilized.

\indent
Charge-exchange (CE) reactions on stable targets at intermediate energies have already been used extensively in nuclear-structure studies as a sensitive probe of the spin-isospin response of nuclei. The study of the isovector giant resonances of unstable nuclei is now gaining interest as a challenging field of research, although with radioactive beams, we often have many orders of magnitude less beam current and much worse energy resolution. 

\indent
Using inverse kinematics, the kinetic energy of the emitted neutrons is relatively small (0.1 - 10 MeV), and with a short flight path (1-2 m), it is possible to obtain an acceptable time of flight (ToF) and energy resolution. The detection of slow neutrons with good efficiency, in addition to the measurement of their energy and angular distribution, requires specially designed spectrometers.

\indent
Such neutron spectrometers have been built by Beyer et al. \cite{beyer} and by Perdikakis et al. \cite{perd} and have been used successfully to study the strength distribution of the Gamow-Teller giant resonance \cite{lenda2}. 
In the framework of the EXL (EXotic nuclei studied in Light-ion induced reactions at the NESR storage ring) and R$^{3}$B (Reactions with Relativistic Radioactive Beams) collaborations, the development of a new spectrometer (the Low Energy Neutron Array (LENA)) was begun in MTA-ATOMKI, Debrecen, as early as 2004. The first article about the detector, which focused mostly on simulations and on some experimental tests with the first prototype detector bars, was published in 2011 by Langer et al. \cite{Lena}. 
Meanwhile, the results of similar developmental efforts (the Wide-angle Inverse-kinematics Neutron Detectors for SHARAQ (WINDS) detector system) at RIKEN and the University of Tokyo have been published \cite{uesaka} and even used successfully in radioactive ion beams (RIBs) \cite{yako}.

\indent
The present work is devoted to presenting the details of the special wrapping method that we used for the final European Low-Energy Neutron Spectrometer (ELENS) and detailed experimental results obtained using the spectrometer. We report on the design and construction of a ToF setup that is larger and has different properties than the LENA detector; it has been tested using various neutron sources and has already been successfully used for in-beam experiments using both stable and gas-jet targets. Recently, an experiment (S408) has been performed at the GSI in Darmstadt to study the absolute, model-independent neutron-skin thickness of the $^{124}$Sn isotope \cite{sn} using the $(p,n)$ reaction in inverse kinematics, using 600 MeV/nucleon relativistic heavy-ion beams and 2-5 mm thick (CH$_{2}$)$_{n}$ and 2 mm thick C targets to constrain the symmetry energy of the equation of state (EoS). The ELENS array was also used in the first in-ring experiments dedicated to nuclear structure at the Experimental Storage Ring (ESR) at GSI \cite{mirko}.

\section{Physical and technical requirements for the design of the ELENS detector}

\begin{figure}[ht]
\centering
\includegraphics[width=90mm]{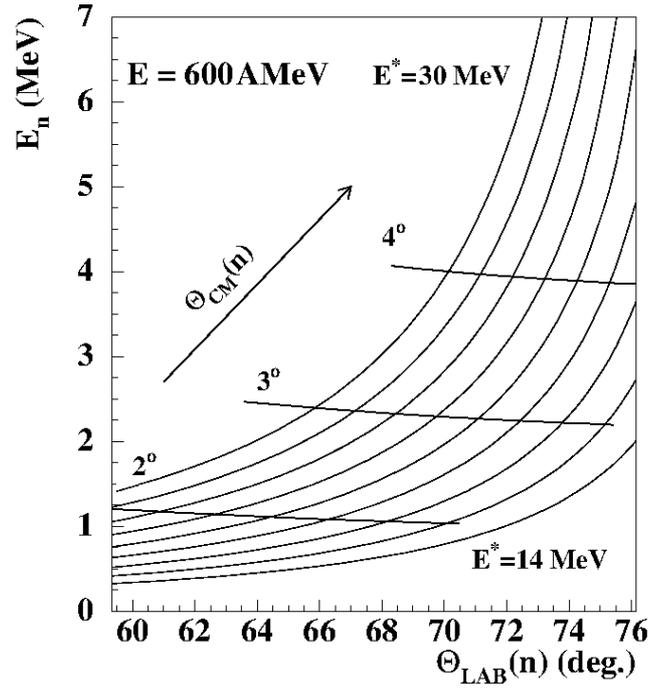}
\caption{Kinematical diagram for the p($^{124}$Sn,$^{124}$Sb)n reaction at a beam energy of 600 MeV/nucleon. The neutron energy is given on the vertical axis, while the neutron angle in the laboratory frame is given on the horizontal axis. The rising solid lines correspond to different excitation energies of the Sb residual; the highest curve corresponds to E$^{\ast}$=30 MeV, while the lowest corresponds to E$^{\ast}$=14 MeV. The straight lines labeled from $2^{\circ}$ up to $4^{\circ}$ represent the center-of-mass angles.}
\end{figure}

\indent
The primary goals of the detector system are to study $(p,n)$-type reactions using radioactive beams and to measure strength distributions as a function of the excitation energy by measuring neutron energies.
The angular resolution plays an important role. In inverse kinematics, the neutron energy depends strongly on the laboratory angle: $\delta \theta = 1^{\circ}$ approximately implies $\delta$E$^{\ast}$=1 MeV, as shown in Figure 1.

\indent
Neutrons do not interact directly with the electrons in matter; the mechanisms for detecting them are based on indirect methods. Neutrons can be scattered by a nucleus, thereby transferring some of their kinetic energy to the nucleus. If enough energy is transferred, then the recoiling nucleus ionizes the material surrounding the point of interaction. For low-energy neutrons, this mechanism is more effective with light nuclei, as the energy transfer is the largest in this case. After the interaction, the recoiled nuclei initiate the release of charged particles, thereby producing light in scintillator material \cite{konyv}. Such neutron detectors are sensitive not only to neutrons but also to gamma rays and cosmic rays, which therefore behave as a source of background. In this work, the suppression of such background radiation will be performed using the ToF method and by gating on the correct energy detected by the scintillators. As an example, the energy loss of cosmic-ray muons in plastic scintillator is approximately 6-8 MeV for muons in the momentum range of 1-100 GeV/c \cite{muon}. These energies are above the energy region investigated by ELENS.

\indent
To construct solid scintillators for neutron detection, liquid scintillators or gas-filled detection media can be used. Gas-filled detectors have very low efficiency in the energy region in which we are interested. Unfortunately, in most international laboratories, the use of liquid scintillators is restricted because of their toxicity. Therefore, plastic scintillators were chosen for neutron detection in the present work; they also have the advantages of fast response, modest cost and easier portability. Plastic scintillator arrays have already been employed successfully in radioactive beams using inverse kinematics to detect low-energy neutrons \cite{lenda2}.

\indent
The following technical requirements were taken into account in the course of designing the detector:
\begin{itemize}
\item{to avoid neutron scattering, it is necessary to minimize the amount of material in the support structures for the detector array and in the surrounding area;}
\item{the positions of the detector frames must be adjustable in the laboratory-angle range of $40^{\circ}$$<$$\theta$$_{lab}$$<$$85^{\circ}$;}

\item{good time resolution is required for the detector (less than 1 ns);}

\item{low cross-scattering of neutrons is necessary;}

\item{and good light collection is required. To increase light collection, it is desirable to surround the detector (i.e., wrap the scintillator bars) with good reflectors that are efficient over a wide range of wavelengths and incidence angles.}
\end{itemize}
\section{Construction of ELENS}

\begin{figure}[ht]
\centering
\includegraphics[width=90mm]{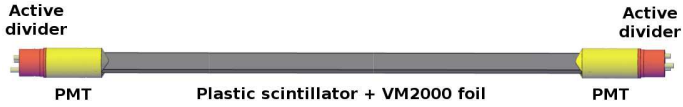}
\caption{A sketch of one detector bar.}
\end{figure}

\indent
The ELENS array consists of 16 single plastic scintillator bars, each of which has a photomultiplier tube (PMT) mounted on each end. A sketch of one bar is shown in Figure 2. The 16 detector bars are arranged in modules; depending on the geometric arrangements, these modules can contain 5 or 11 scintillator bars. The array is designed to measure neutron energies using the time-of-flight  technique in the kinetic-energy region from a few hundred keV to a few MeV. Each bar consists of fast plastic scintillator material with dimensions of $10 \times 45 \times 1000$ mm$^3$. The 1000 mm length corresponds to a large opening angle, and the 45 mm side width provides a high probability of interaction between the neutrons and the scintillator material but remains small enough for precise position determination. The 10 mm profile width provides an angular resolution of approximately $1^{\circ}$ if the source is approximately 1 m away from the array.

\indent
The type of plastic used for the bars is UPS89, and it was delivered by the Amcrys-H company in Ukraine. Its properties (light output = 65 $\%$ of anthracene, wavelength of maximum emission = 418 nm, rise time = 0.9 ns, decay time = 2.4 ns, H/C atomic ratio = 1.104 and light attenuation length = 360 cm \cite{ukr}) are similar to those of NE102A. These scintillators have a polystyrene matrix for the detection of gamma radiation and fast neutrons. High-viscosity silicon grease was used for the light coupling (type: EJ-560 Optical Interface sheet). To avoid neutron scattering, careful construction of the holder for the detector array was very important. It should contain as little material as possible, but it should be stable and rigid. ITEM Profile “6” 30 $\times$ 30 (0.0.419.01) anodized natural aluminum components were used for the detector-array holder \cite{item}. 
\begin{figure}[ht]
\centering
\includegraphics[width=80mm]{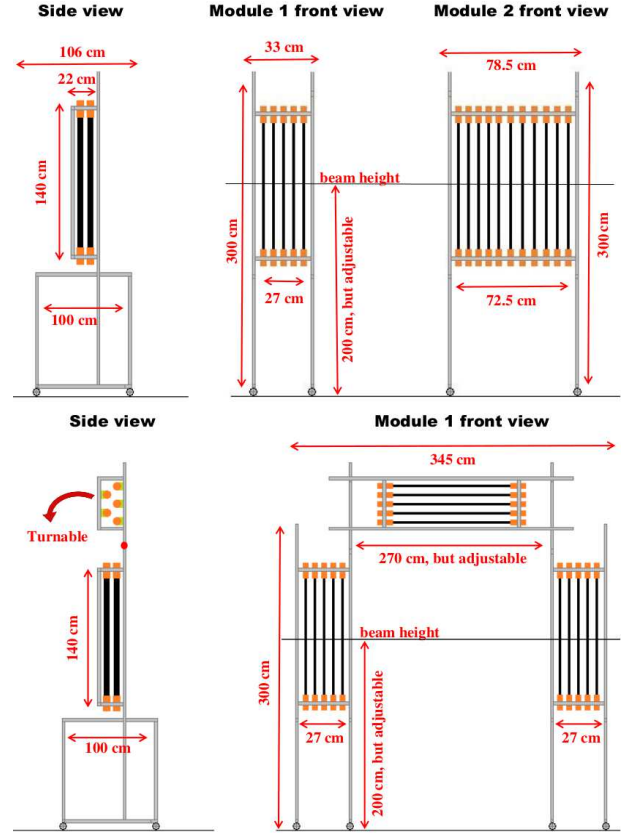}
\caption{The upper panel shows the first type of geometry, which has 2 modules. The detector-array holder can be placed on either side of the beam line to study the selected angular region. The larger module (11 bars) can be used to cover a larger angular range, while the smaller one can be placed within the real region of interest that one would like to study with better statistics. The bars are placed in two parallel planes. The detectors in the first plane are shifted by 3.75 cm with respect to those in the second to obtain uniform angular coverage. The second type of geometry (lower panel) has 3 modules, each of which contains 5 bars. The detector-array holder can be placed surrounding the beam line with each module in the same angular position. This geometry yields a higher detection efficiency in the selected solid-angle range.}
\end{figure}

\indent
To determine the positions of the hits and to reduce the background, each bar is coupled to two photomultiplier tubes (PMTs). Two different types of 51 mm diameter photomultiplier tubes were tested to achieve the lowest possible detection threshold: Hamamatsu R2059 and Photonis XP2262. There was no significant difference observed between the two tubes. Ultimately, the Photonis tubes were chosen because the quartz window of the Hamamatsu was not necessary in our wavelength range. The Photonis PMT features a 12-stage amplification section that provides a maximum gain of $3 \times 10^7$. The lime-glass window material and the maximum sensitivity of the fast tube (rise time: 2.3 ns) at a 420 nm wavelength are ideally suited to the UPS89 plastic scintillators \cite{phot}. By combining the time and pulse-height information from the PMTs, the timing of a neutron hit, the corresponding scintillation-light output, and the neutron-hit position along the longest side of each ELENS bar can be determined.

\begin{figure}[ht]
\centering
\includegraphics[width=75mm]{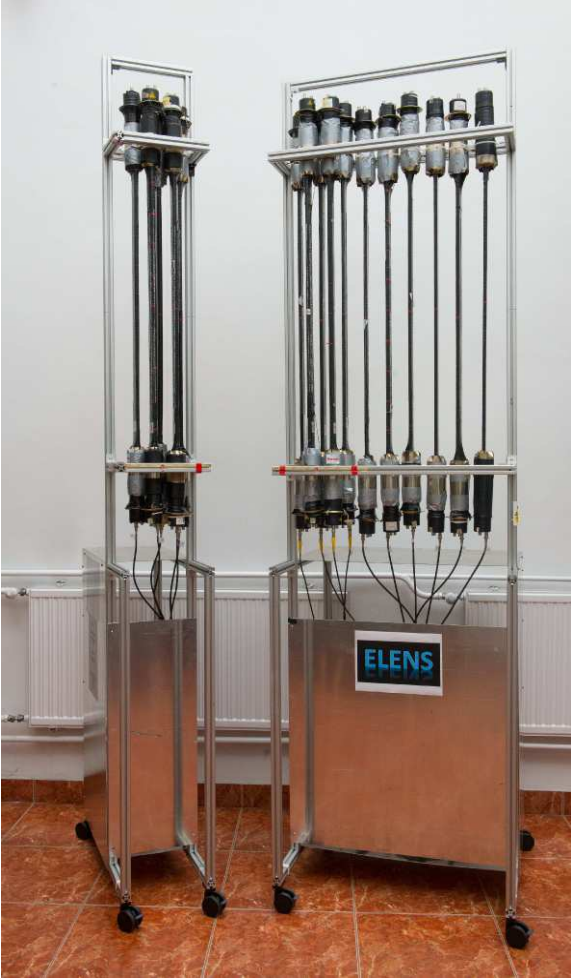}
\caption{A photograph of the ELENS array. The PMTs are covered with special mu-metal sheets against possible interference effects caused by the magnetic fields.}
\end{figure}

\indent
A variable geometry was developed for ELENS. The schematic layouts of the two typical configurations of the detector system are shown in Figure 3. There are two main configurations, which are optimized for charge-exchange $(p,n)$ experiments in inverse kinematics. In Figure 3, the upper panel shows the first type of geometry, which has 2 modules (one containing 5 bars and one containing 11 bars). The bars are placed in two parallel planes. The detectors in the first plane are shifted by 3.75 cm with respect to those in the second to achieve uniform angular coverage at 1 m from the target. In this case, the angular separation between two bars in the same plane is 3.97$^{\circ}$ (in the case of bars in each of the two parallel planes, the angular separation is 1.98$^{\circ}$). Geometry 1, with its 11-bar and 5-bar modules, is suitable for studying giant resonances over a wide energy range. The larger module (11 bars) can be used to cover a larger angular range (background), while the smaller one can be placed within the real region of interest (effect) to allow it to be studied with better statistics. The detector-array holder can be placed on either side of the beam line, and the array covers scattering-angle ranges of approximately 8 (smaller module) and 20 (larger module) degrees in the laboratory frame.

\indent
The second type of geometry (lower panel) has 3 modules, which contain 5 bars each. In the case of two parallel planes with a half-step shift, the angular separation is 1.98$^{\circ}$. The detector-array holder can be placed surrounding the beam line, and ELENS covers a 8$^{\circ}$ range of the scattering angle if the angular position of each module is the same. Alternately, the ELENS array can cover a maximum scattering-angle range of 24$^{\circ}$ if each of the 3 modules covers a different 8$^{\circ}$ wide angular range.
A photograph of the ELENS system (in the geometry 1 arrangement) is shown in Figure 4.

\section{The wrapping procedure}
\begin{figure}[ht]
\centering
\includegraphics[width=70mm]{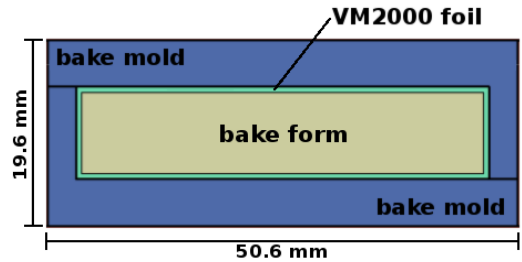}
\caption{The schematic cross-sectional view of the bake molds and, between them, the bake form. The VM2000 multilayer reflector foil placed in the baking set is also shown.}
\end{figure}

\indent
The response of the plastic scintillator is very non-linear for low-energy protons, which are created via the elastic scattering of the neutrons \cite{brinks}. The electron-equivalent light creation of protons (E$^{p}_{ee}$) depends non-linearly on the energy of the protons: E$^{p}_{ee}$ = 0.16 $*$ (E$^{p})^{3/2}$ \cite{brinks}. As a rule of thumb, 10 keV electrons and 0.2 MeV protons yield approximately similar amounts of light in a typical plastic scintillator. The detection of such weak light pulses is limited by the single-electron noise (SEN) of the photomultiplier tubes. By requiring coincidences between the two phototubes connected to the scintillator, the effect of SEN can be reduced \cite{Lena}, but because of the light attenuation along the scintillator bars, the coincidence efficiency is also a limiting factor for low-energy particles.

\indent
Very good light-collection efficiency is required for the detection of these small signals, so a proper wrapping material and the tight fitting of the foil onto the plastic were important criteria for ensuring a sufficiently high-quality light connection. There are several possible wrapping-material candidates for plastic scintillators. The most popular materials are 3M Radiant Mirror Films, polytetrafluoroethylene (PTFE) tapes, plastic tapes, gold-coated tapes, aluminum foils and white diffuser paint or tape. The most important properties of materials for this application are their high reflection rates. In addition, these materials must be mechanically stable and flexible within certain limits. After several tests, it was decided that the detector bars should be wrapped with a specially treated VM2000 multilayer reflector foil, which has recently been produced by 3M. Several studies have been performed previously to study the effect of wrapping scintillator bars with VM2000 \cite{vm}, \cite{kalinka1}, \cite{kalinka2}, which is a multilayer reflective foil based on a novel technology \cite{we}. This foil has a good reflection coefficient of $R > 97$$\%$ for $\lambda \geq 400$ nm and R $= (98.5 \pm 0.3)$$\%$ at 430 nm \cite{mott}. 

\indent
Because of the thickness and rigidity of the multilayer material, it is very difficult to fold, wrap or create any kind of a crease in VM2000 foil without decreasing its light-guidance parameters. There are many methods of achieving optimal wrapping (with VM2000), but all these methods use some mechanical interactions with the foil. VM2000 is composed of several hundred different layers, so it is sensitive to such external mechanical influences as cuts and abrasions. Any abrasion or cut may compromise the optical properties of the foil. Along the cut edges, the thickness of the material changes (becomes thinner), or microscopic cracks are generated, thereby reducing the reflectance. To ensure the proper fitting of the reflective wrapping foils to the scintillator bars, the foils were formed using a special heat treatment prior to the actual wrapping process. A special baking set was fabricated, and a heating and cooling cycle was performed with the foil loaded into the set.

\indent
As the first step, we prepared a special baking set with 2 bake molds and, between them, a bake form (see Figure 5). The material of the baking set was aluminum, and its surface was polished. The size of the form was smaller than the original plastic by 0.5 mm in the two smaller dimensions, i.e., $1000 \times 44.5 \times 9.5$ mm$^3$. There are two reasons for the smaller size of the bake form:
\begin{itemize}
\item{to achieve the optimal light contact between the two surfaces, the foil must be stretched on the scintillator bar, and} 
\item{it was necessary to take the thermal expansion of the bake form during the baking process into account.}
\end{itemize}
A slice of VM2000 foil was cleaned. The VM2000 foil is covered with a protective film (as provided from the factory) that protects against external damage. The surface of the protective film was gently cleaned before the treatment to remove any contamination that could affect the smoothness of the foil. After the cleaning, the foil was laid between the molds and the form. This must be a very precise process, as any stretching, stress or shrinkage of the reflective film can compromise the optical properties of the foil. After baking, the protective film was removed from the VM2000 foil. The use of a dust-free room was also important to minimize the risk of contamination.

\begin{figure}[ht]
\centering
\includegraphics[width=90mm]{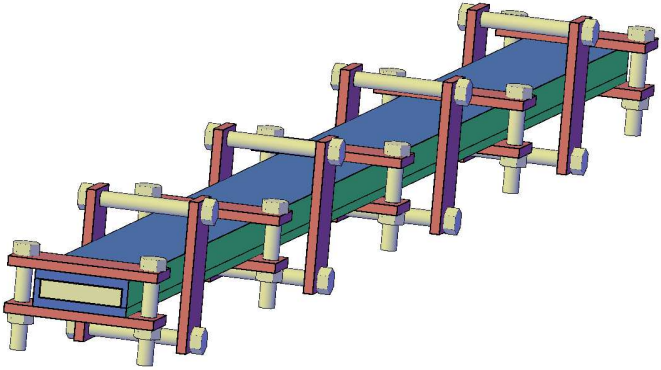}
\caption{The schematic layout of the baking set with the iron clamps. The bake molds are fixed in place from the outside with 9 iron clamps.}
\end{figure}

\indent
Once the foil had covered the aluminum bake form with the aid of the exterior aluminum mold fixture, then we fixed it in place from outside with iron clamps (see Figure 6). The distance between the clamps was 10 cm. The wrapping mold with the foil inside was placed into an electric oven, and it was baked for a duration of 2 hours at 115 $^{\circ}$C. After the mold was removed from the oven, it was cooled for 24 hours. After the above treatment, the formed foil was stored for one week before the wrapping was performed. The optical parameters of the foil were not degraded following this procedure; indeed, it performed better than before the treatment (see section $5.4.$).

\indent
The scintillator bars have been wrapped with one layer of treated foil, a layer of aluminum foil and finally black insulating tape to ensure both light-tightness and proper light propagation through the bar.

\section{Response of the ELENS detector}

\subsection{Simulations of the response of plastic scintillators}

\indent
A simulation of the response of the selected scintillator to neutrons was performed using the GEANT4 toolkit \cite{geant}. GEANT4 simulates neutron transport for thermal energies up to 20 MeV.
Elastic and inelastic scattering and neutron capture and fission are treated by referring to the G4NDL3.13 cross-section data. The detection of neutrons is based on their interaction with the scintillator material. The following processes by which neutrons interact with the detector material were taken into account:
\begin{itemize}
\item{elastic scattering on hydrogen and $^{12}$C,}

\item{inelastic scattering on $^{12}$C, and}

\item{$^{12}$C(n,$\alpha$) and $^{12}$C(n,p) reactions.}
\end{itemize}
With each interaction, the neutron deposits a portion of its initial energy, until it escapes from the detector or is captured. The energy that is deposited by the neutron in the detector volume is converted into light output, depending on the type of interaction. The electron-equivalent energy ($E_{ee}$) of the deposited energy for protons and $\alpha$ particles can be calculated using the following empirical expression:
\begin{equation}
${$E_{ee}$=a$_1$E$_p$$-$a$_2$[1.0$-$exp($-$a$_3$$*$E$_p^{a_4}$)]}$
\end{equation}
where the electron energy ($E_{ee}$) and the proton (E$_p$) or alpha energy (E$_{\alpha}$) are in units of MeV. The values of the parameters a$_1$-a$_4$ have been provided by Cecil et al. \cite{cecil} for protons and alphas as well.
In the case of scattering on C, the light output is very small. The electron-equivalent energy was calculated using the following relation:
\begin{equation}
${E$_{ee}$=c$*$E$_C$}$
\end{equation} 
where c=0.02 MeVee/MeV \cite{ds}. For each scattering or nuclear reaction of a neutron within the scintillator, the proper light output was calculated and stored. The summed light output of all interactions of the neutron was taken to be the light output of the scintillator for one neutron event.

\indent
The neutron interaction was studied in an energy range of 200 keV - 10 MeV under the assumptions that the scintillator material contained 100\% carbon or 100\% hydrogen. The simulations demonstrated that the dominant process is the elastic scattering on H and C. The light output originating from the neutron-carbon interaction is very small; it is approximately 1\% at a neutron energy of 6 MeV. Its contribution to the efficiency is negligible for neutron energies below 8 MeV. For neutron energies below 8 MeV, the dominant detection mechanism is the proton-neutron elastic scattering.

\subsection{Experimental study of the response of the array}
The response of the ELENS detectors to monoenergetic neutrons was investigated at the Physikalisch-Technische Bundesanstalt (PTB) accelerator facility in Braunschweig, Germany. The experiments were performed in a large ($24 \times 30$ m$^2$), temperature-controlled experimental hall with a height of 14 m. To reduce the neutron scattering, an intermediate floor of gridded aluminum was placed 4.5 m above the ground. Quasi-monoenergetic neutrons were produced with different energies, and the effect of the wrapping on the efficiency of the detector was studied. The distance of the ELENS detector from the target was approximately 5 m (from the target to the front side of the scintillators). The ELENS bars were mounted at $42.5^{\circ}$. Table 1 presents the reactions used, the proton-beam energies, the target parameters and the obtained kinetic energies of the neutrons (at $42.5^{\circ}$). For the measurements, the proton beams were produced by a 3.5 MV Van de Graaff accelerator. The measurements were performed at four different neutron energies. The energy values of the neutrons produced in the target were calculated using the EnergySet program \cite{eset}. The program includes the geometry of the target and all relevant reaction cross sections. Using dedicated quasi-monoenergetic neutrons at kinetic energies of 240, 471, 925 and 2014 keV, the neutron-detection efficiency was also determined, as will be described below.

\begin{table}
\caption{Nuclear reactions used to create quasi-monoenergetic neutrons at PTB, Germany.}
\centering
\begin{tabular}{cccc}
\hline
Reaction & E$_p$ /keV/ & Target & E$_n$ /keV/ \\
\hline
$^7$Li(p,n)$^7$Be & E$_p$=2.03 MeV & LiF on Ag backing & 240 \\
$^7$Li(p,n)$^7$Be & E$_p$=2.3 MeV   & LiF  on Ag backing & 471\\
T(p,n)$^3$He & E$_p$=2.1 MeV   & T/Ti  on Ag backing & 925 \\
T(p,n)$^3$He & E$_p$=3.36 MeV &  T/Ti on Ag backing & 2014 \\
\hline
\end{tabular}
\end{table}

\subsection{Electronics}
\begin{figure}[ht]
\centering
\includegraphics[width=89mm]{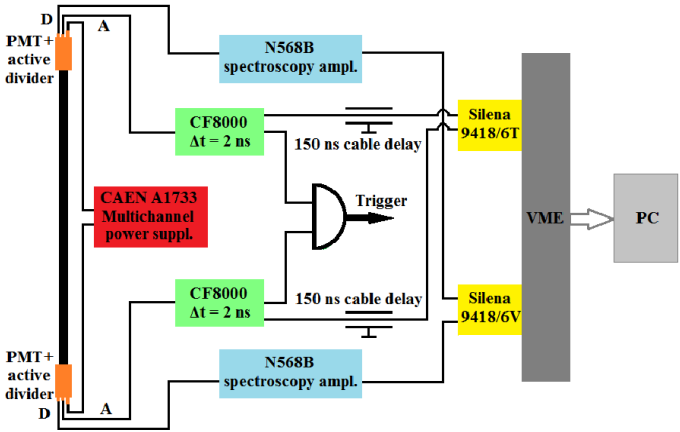}
\caption{Schematic drawing of the electronics used.}
\end{figure}
\indent
We used VD122K/B active dividers, which ensured the high-count-rate capability of the system. The PMTs constructed by Photonis and CAEN A1733-type 12-channel negative power supplies ensured the long-term stability of the system. The amplification of the photomultiplier tubes depends strongly on the stability of the output voltage of the power supplies. The output voltage from a CAEN A1733 has an accuracy of $\pm$ 0.5 V $\pm$ 0.3\% of the reading \cite{caen}. The anode signals were connected to CF8000 8-channel constant-fraction discriminators with 2 ns internal delays. The delayed (150 ns cable delay) signals were then connected to 32-channel VME (Silena 9418/6T) time-to-digital converters (TDCs). The energy signals were acquired from the last dynode of the PMTs. They were slowed down to have decay-time constants of 2.5 $\mu$s and were fed into CAEN N568B 16-channel spectroscopy amplifiers. The output signals were then digitized with 32-channel VME (Silena 9418/6V) ADCs. The VME modules were read out via a Wiener A32/D32 VME bus controller \cite{prog1}. High-performance data-acquisition software \cite{prog2} was used for storing the data and for the online monitoring of the spectra (see Figure 7). The data were recorded in event mode and later analyzed in more detail offline.

\subsection{Test of the wrapping}
\indent
In the ToF method, the efficiency of the detection depends on the amount of light collected at the surface of the PMT and the gain of the PMT, and the thresholds of the constant-fraction discriminators also play a significant role. This dependence provided us with a good opportunity to measure the effectiveness of the light-collection by measuring the detection efficiency for neutrons. The gains of the PMTs were tuned to be equal using a $^{60}$Co source. The thresholds of the CFDs were also set to be equal. (The light attenuation modified by the wrapping was also tested using a radioactive source; see section $5.8.$ below.) 

\indent
As a first step, we modeled two different materials, Al and Teflon foils, for wrapping the scintillators in our simulation. The effect of 1 mm thick wrapping materials on the efficiency is negligible (less than 0.5$\%$). Three types of wrapping were compared experimentally. 

\indent
In Table 2, the light-collection efficiencies of the bars wrapped with Teflon (+ Al foil + black plastic) and those wrapped with specially treated (as explained in the previous section) VM2000 foil (+ Al foil + black plastic) are compared to bars that were wrapped with VM2000 foil without baking (+ Al foil + black plastic). The values were normalized to the data collected using the untreated VM2000 foil. We concluded that using the specially treated VM2000 foil (referred to as VM2000* in Table 2), we can gain approximately 15-20$\%$ in relative neutron-detection efficiency with respect to the untreated foil.

\begin{table}
\caption{The relative light-collection efficiencies of the detector bars at various energies and with various wrappings. The VM2000* notation represents the specially treated VM2000 foil. The uncertainties are less than 2\%.}
\centering
\begin{tabular}{ccccc}
\hline
& \multicolumn{4}{c}{Neutron energy} \\
\hline
Wrapping & 210 keV & 471 keV & 925 keV & 2014 keV \\
\hline
Teflon tape & 95$\%$ & 96$\%$ & 96$\%$ & 98$\%$ \\
VM2000 & 100$\%$ & 100$\%$ & 100$\%$ & 100$\%$ \\
VM2000* & 116$\%$ & 115$\%$ & 118$\%$ & 120$\%$ \\
\hline
\end{tabular}
\end{table}

\subsection{Neutron scattering between the scintillator bars}
\indent
The planned applications of ELENS are sensitive to the exact positions of the primary interaction events of the neutrons with the plastic, so scattering between the detectors can be a serious problem. Monte-Carlo simulations were performed to study the scattering between the detectors using a 925 keV neutron source. The experimental results were compared to the results of the simulations. As shown in Figure 8, there is a fair agreement ($<$1\%) between the values for scattered neutrons with 925 keV kinetic energies obtained from the Monte-Carlo simulated data and the experimental data. The data from these simulations were also used to determine the distances between the middle planes of the bars of the spectrometer. We attempted to find the optimal distances to achieve both sufficient angular resolution (see section $3.$) and scattering across the bars of less than 10\%. The optimal calculated distance was found to be 7.3 - 7.8 cm. The distance between the middle planes of the bars of the spectrometer is 7.5 cm; this is the same for all configurations. 

\indent
To obtain experimental information concerning the scattering between the detector elements, a series of experiments were performed at PTB, Germany. In these measurements, the scintillator bars were arranged in two parallel rows, as shown in Figure 8. The lower left bar was irradiated with neutrons (we required that detector to be fired in the off-line analysis in coincidence with the other detectors) and the ratios of the scattered neutrons detected in the other bars were measured. The scattering was studied experimentally at neutron energies of 471 keV, 925 keV and 2014 keV. The probabilities of scattering from the irradiated scintillator to the others are shown in Figure 8. We note an insignificant cross-scattering probabilities ($<$5\%) between the detector bars for the scattered neutrons with a kinetic energy of 925 keV obtained from the experimental data (panel a)) and the data from the Monte-Carlo simulations (panel b)).
The scattering probabilities at the other two energies are shown in the c) and d) panels of Figure 8. 
\begin{figure}[ht]
\includegraphics[width=89mm]{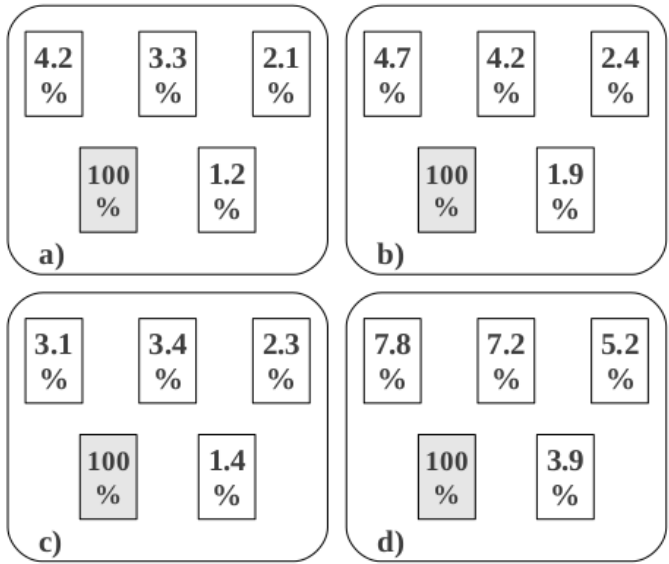}
\caption{Panels a) and b) represent the experimental and simulated cross-scattering probabilities for 925 keV, while panels c) and d) represent the experimental data measured at 471 keV and 2014 keV, respectively. The relative uncertainties are less than 5\% for 925 keV and 471 keV and less than 3\% for 2014 keV.}
\end{figure}

\subsection{The efficiency of ELENS}
\indent
The detection efficiency of the spectrometer was first measured using a thin $^{252}$Cf fission source, for which the neutron spectrum is precisely known \cite{mead} and can be approximated by a Maxwellian distribution, 
\begin{equation}
${N(E) = E$^{1/2}$ $*$ exp(-E/1.565)}$
\end{equation}
in the energy range of 0.5$<$E$<$10.0 MeV, where E is given in MeV.
\begin{figure}[ht]
\centering
\includegraphics[width=90mm]{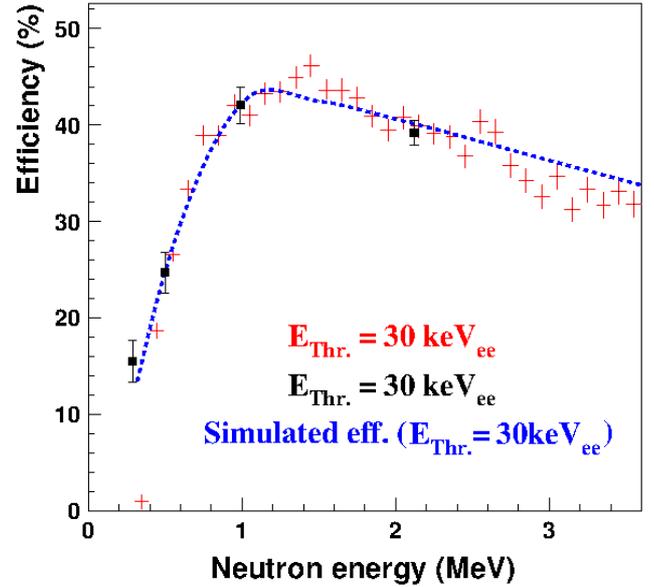}
\caption{The efficiency of ELENS in the 0.25 MeV to 3.5 MeV region. The points marked with crosses represent the experimental results of $^{252}$Cf measurements with a 30 keV$_{ee}$ threshold, which is close to the single-electron noise (SEN), while the black squares represent the efficiency measurements at PTB using monoenergetic neutrons. The dashed line corresponds to the simulated efficiency.}
\end{figure}

\indent
The kinetic energies of the neutrons were determined using the ToF method.
The ToF start signal was generated by the fission fragments detected using a thin (0.2 mm) plastic scintillator glued to the surface of a XP2262 PMT tube, which was equipped with a VD122K/B active divider. The fission source mounted on the start detector was placed 100 cm from the neutron detectors. The anode signal of the active divider was connected to a CF8000 discriminator channel with a 2 ns delay. To suppress the random coincidences caused by the alpha particles, the threshold of the discriminator was set just above 6.1 MeV, corresponding to the signals produced by the intense alpha particles from the primary $^{252}$Cf alpha-decay branch. The threshold for the neutron detection was set to 30 keVee on each PMT. The results of the measurement are plotted in Figure 9. 

\indent
The neutron-detection efficiency of each ELENS bar was also measured individually at PTB, Germany. The ``geometry 1’’ configuration (see Figure 2, upper panel) was used for the determination of the efficiency of the system. A pulsed proton beam (f$_{\circ}$=5 MHz, $\delta$t$\leq$1 ns) was used to produce neutrons with well-defined timing. The measurement of the ToF with respect to the RF signal provided suitable identification of the neutrons. A typical ToF spectrum is shown in Figure 10.
\begin{figure}[ht]
\centering
\includegraphics[width=90mm]{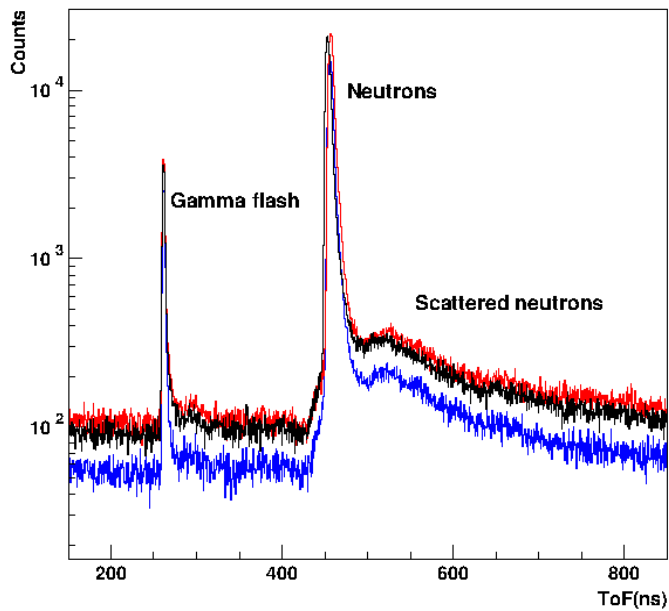}
\caption{Measured time-of-flight spectra for 3 different detector bars at a neutron energy of 2014 keV. The first two (higher) represent the same number of scattered neutrons, while the third bar (lower) was positioned at the side of the array, so it detected fewer scattered neutrons. This difference in positioning could account for the lower background measured for the third bar.}
\end{figure}
\begin{figure}[ht]
\centering
\includegraphics[width=90mm]{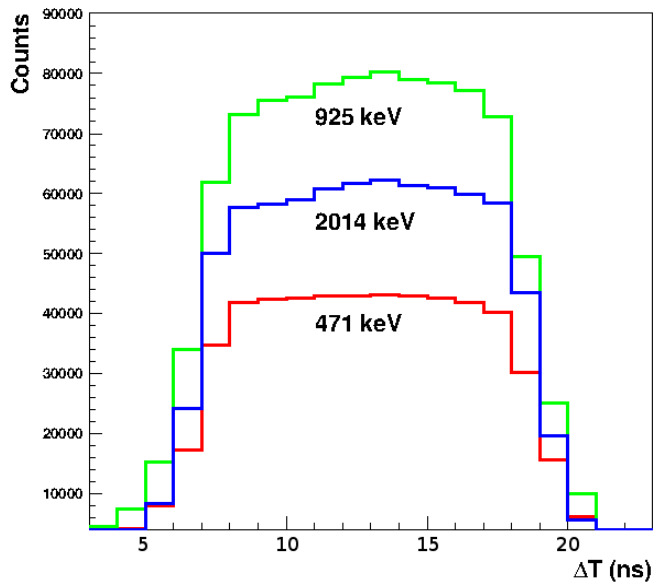}
\caption{The time-difference distribution between the 2 PMTs of a detector bar, which is proportional to the relative efficiency along the detector as a function of the time difference, measured at three different neutron energies (471 keV, 925 keV, 2014 keV). The threshold for the neutron detection was set to 30 keVee on each PMT. The statistical errors can be calculated from the count$/$channel values.}
\end{figure}

\indent
Two standard BF$_3$ long neutron counters were mounted as reference counters at $98^{\circ}$ and $16^{\circ}$ with respect to the direction of the proton beam. The ELENS bars were mounted at $42.5^{\circ}$. The detector bars with different parameters (wrappings) were calibrated, and the wrappings with Teflon tape, simple VM2000 foil and treated VM2000 foil were compared. The treated VM2000 exhibited the best efficiency. These values are plotted in Figure 9. 

\indent
The efficiency of ELENS was simulated using the GEANT4 Monte-Carlo (MC) code. In the simulations, we assumed a point source of neutrons positioned 1 meter from the center of the scintillator bar. The efficiency values calculated using the GEANT4 MC simulations (dashed line) are compared to the experimental efficiencies in Figure 9. The simulated and experimental results are in reasonably good agreement (typically within $\pm 10-16\%$). The efficiency of the detector along the scintillator bars is constant within $\pm 5-7\%$, as shown in Figure 11.

\subsection{Time and position resolution of the detector system}
\indent
The position resolution was measured using a $^{90}$Sr electron source with a continuous distribution of electron 
energies. The emitted electrons were collimated to a 5 mm diameter spot on the surface of the bar.

\indent
Figure 12 shows the time differences between the PMTs of a single detector, which were measured for interactions at 5 different positions with respect to the middle of the bar in 10 cm steps. To obtain the position of a detected event along the detector, the time difference between the signals registered by the two photomultipliers was used. Only the events in which both tubes fired were taken into account in the analysis. On average, the effective length (sensitive length) \cite{Lena} of the detector bars of the ELENS spectrometer has been determined to be $(98.5 \pm 0.3)$ cm.
\begin{figure}[ht]
\centering
\includegraphics[width=89mm]{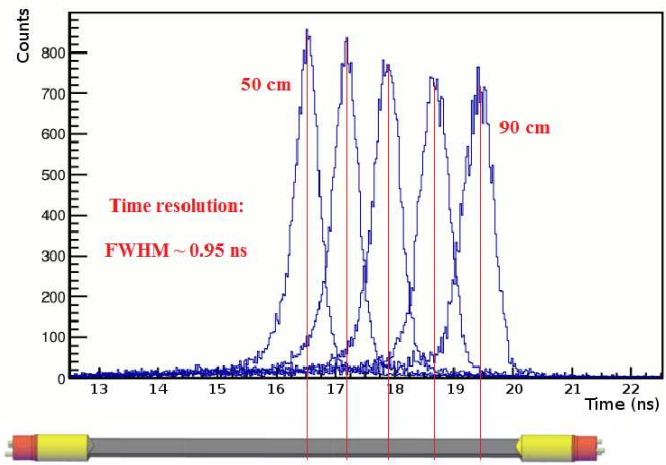}
\caption{Time-difference distributions between the two PMTs of a single detector. The collimated $^{90}$Sr electron source was moved along the detector in steps of 10 cm. The peak areas are normalized to the peak area measured at the center of one detector bar of ELENS.}
\end{figure}

\indent
The time resolution was deduced for each bar. As it depends on the position of the detection, the detector-bar time resolution was defined as the width (FWHM) of the time-distribution peak in the center of the bar (see Figure 12). The time resolution was found to vary between 680 ps and 950 ps (FWHM) for the different detector bars. The average time resolution of the 16 bars is 840 ps. 

\indent
The position resolution was found to be 7.2 - 8.3 cm (FWHM). The average position resolution of the 16 bars is 7.9 cm. Position resolutions are usually worse near the ends of a detector \cite{hasi}. In our case, at the two ends of the detector bars, the resolution degrades by approximately 50$\%$.

\subsection{Light attenuation in the scintillator}

\begin{figure}[ht]
\centering
\includegraphics[width=90mm]{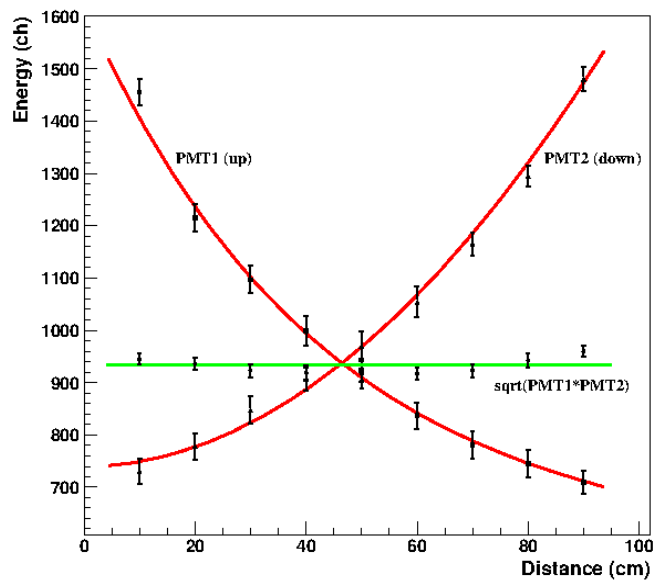}
\caption{Dependence of the signal amplitude on the distance between the $\gamma$-ray interaction and the surfaces of the PMTs. Measurements with a $^{60}$Co source were performed for a single bar. All results were calculated from the original data.}
\end{figure}
\indent
One of the major characteristics that describes the performance of a plastic scintillator is the light-attenuation length. The light-attenuation length of a plastic scintillator bar is defined as the length traveled by the light signal over which the magnitude of the signal is reduced by a factor of $e$, and it depends upon the bulk transmission of the scintillator, its thickness and its shape.
The light-attenuation length was derived from the dependence of the pulse heights of the PMT outputs on the position of the $^{60}$Co source, and it therefore includes the effects of the scintillator surface \cite{assa}.
A measurement was performed to study the effective attenuation length of the bars of ELENS by moving a $^{60}$Co source along the bar. Starting from one end of the bar, the source was moved in 10 steps. The observed light-attenuation length \cite{assa} is $(128.2 \pm 1.7)$ cm. This means that light can travel relatively long distances in the scintillator material without significant attenuation. As shown in Figure 13, the light attenuation changes exponentially as a function of the distance of the source from the PMT, which is expected behavior \cite{brinks}, \cite{kurata}. The light attenuation of the detector material with the specially treated VM2000 reflecting foil is approximately 50$\%$.

\indent
In the case of $^{60}$Co, it was necessary to use the Compton edge to characterize the signal amplitude. We estimated the position of the Compton edge to be at 70$\%$ of the height of the Compton slope. The addition of reflective wrapping (treated foil) decreases the rate at which light is lost during transport through the length of a scintillator bar. The advantage of using such wrapping is also reflected by the fact that our measured speed of the light inside the scintillator bar is $(13.3 \pm 0.4)$ cm/ns, which is much smaller than the speed of light in the scintillator material supporting the multiple reflection inside the detector bar \cite{buta}.

\section{Conclusion}

\indent
The European Low-Energy Neutron Spectrometer (ELENS) was developed for the detection of low-energy neutrons from charge-exchange reactions in inverse kinematics. The spectrometer has an angular resolution of 1$^{\circ}$, a time resolution of 840 ps (averaged over the 16 bars) and a position resolution of 7.2 - 8.3 cm. The light attenuation of the detector material is approximately 50$\%$. The spectrometer is capable of measuring low-energy neutrons with relatively high light yield (compared to the light yield for $\gamma$ rays) and high efficiency; the latter is approximately 40$\%$ for neutrons with a kinetic energy of 1 MeV.

\indent
Based on our experiences during the first experiment conducted with this detector array, the properties demonstrated by the ELENS system confirm that it will be well suited for new inverse-kinematics experiments with exotic beams.

\section{Acknowledgments}
\indent
This research was supported by the European Union and the State of Hungary, co-financed by the European Social Fund in the framework of the T\'AMOP 4.2.4.A/2-11-1-2012-0001 $‘$National Excellence Program$’$ (Nemzeti Kiv\'al\'os\'ag Program – Hazai hallgat\'oi, illetve kutat\'oi szem\'elyi t\'amogat\'ast biztos\'it\'o rendszer).

This work was also supported by the European Community FP7 - Capacities, contract ENSAR No. 262010, the Hungarian OTKA Foundation No. K 106035, the Spanish FPA 2011-24553 and the NUPNET project.

N. K., S. B, M. A. N. and C. R. are financially supported in the context of the contract between the University of Groningen (RuG) and the Helmholtzzentrum f\"ur Schwerionenforschung GmbH (GSI), Darmstadt.







\end{document}